\documentclass[aps,prd,twocolumn,amsmath,amssymb,superscriptaddress]{revtex4-2}

\usepackage[utf8]{inputenc}
\usepackage[T1]{fontenc}
\usepackage{graphicx}
\usepackage[colorlinks=true,allcolors=blue]{hyperref}

\newtheorem{theorem}{Theorem}[section]
\newtheorem{proposition}[theorem]{Proposition}
\newtheorem{lemma}[theorem]{Lemma}
\newtheorem{corollary}[theorem]{Corollary}
\newtheorem{remark}[theorem]{Remark}
\newtheorem{example}[theorem]{Example}

\newenvironment{demonstration}[1][Proof]%
  {\par\medskip\noindent\textit{#1.}\ }%
  {\hfill$\blacksquare$\par\medskip}

\newcommand{\dd}{\mathrm{d}}
\newcommand{\D}{D_t}                       

\begin{document}

\title{Lorentz symmetry and Noether currents for classical\\
higher-derivative systems: towards nonlocal dynamics}

\author{Emmanuel M'po Takou}
\email{emmanuelmpotakou@gmail.com}
\affiliation{Facult\'e des Sciences et Techniques (ICMPA-UNESCO Chair),
Universit\'e d'Abomey-Calavi, 072 BP 50, Benin}

\author{Vincent Lahoche}
\email{vincent.lahoche@cea.fr}
\affiliation{Universit\'e Paris-Saclay, CEA, Palaiseau, F-91120, France}

\author{Dine Ousmane Samary}
\email{dine.ousmanesamary@uac.bj}
\affiliation{Facult\'e des Sciences et Techniques (ICMPA-UNESCO Chair),
Universit\'e d'Abomey-Calavi, 072 BP 50, Benin}

\date{\today}

\begin{abstract}
Nonlocal theories, whose dynamics involves infinitely many derivatives, appear
in string field theory, in ghost-free gravity and on noncommutative spacetime.
For such theories to be physically acceptable, their Poincar\'e charges
(energy, momentum, angular momentum and boost charge) must exist and be
conserved, but it is not clear how to define them, since the canonical
momenta involve derivatives of arbitrarily high order. We address this
question for a relativistic point particle, regarding a nonlocal Lagrangian
as the limit of local Lagrangians containing derivatives up to order $q$.
Using the time $t$ of an inertial observer as evolution parameter, so that
Lorentz boosts act on time itself, we derive the action of the Poincar\'e
group at every order $q$, prove a Noether identity with an explicit
conserved current, and write the ten charges in Ostrogradsky variables. The
higher derivatives generate an internal spin, which makes the particle
oscillate (Zitterbewegung) around a uniformly moving center of energy. In
$1+1$ dimensions we determine all Lorentz-invariant Lagrangians: they depend
only on the proper acceleration and its derivatives. We show that the charges
realize the Poincar\'e algebra, and obtain the mass and the spin in closed
form. For the limit $q\to\infty$, we prove that the charge of each truncated
theory drifts by an amount fixed exactly by the neglected part of the
equations of motion; this gives explicit conditions under which the nonlocal
charges exist and are exactly conserved. In a nonlocal model these conditions
select the physically admissible solutions, and the model admits exact
self-accelerated solutions with zero mass, which we show to be unstable.
\end{abstract}

\maketitle

\section{Introduction}
\label{sec:intro}

\paragraph*{Nonlocality in modern physics.}
Locality, the principle that the dynamics at a spacetime point is determined by
the fields and a finite number of their derivatives at that point, underlies
classical mechanics and relativistic field theory. Several frameworks of
current interest nevertheless relax it. String field theory and $p$-adic
strings lead to actions with infinitely many derivatives
\cite{EliezerWoodard1989,BarnabyKamran2008}; ghost-free and
super-renormalizable theories of gravity rely on entire functions of the
d'Alembertian \cite{Biswas2012,Modesto2012}; and field theories on
noncommutative spacetime are nonlocal through the star product
\cite{SeibergWitten1999,DouglasNekrasov2001}. Field theories with nonlocal
actions go back to the work of Pais and Uhlenbeck \cite{PaisUhlenbeck1950}.

\paragraph*{Relativistic nonlocal dynamics.}
A first requirement on any such theory is compatibility with special
relativity: interactions between distinct events must preserve an appropriate
form of Lorentz covariance, which raises questions about causality, about the
definition of the dynamical variables and about the equations of motion. The
historical prototype is the direct-action electrodynamics of Fokker, Wheeler
and Feynman \cite{Fokker1929,WheelerFeynman1945,WheelerFeynman1949}. There is
no field: two worldlines $x_{1,2}^\mu$ with charges $e_{1,2}$ are coupled
through a Lorentz-invariant kernel supported on lightlike separations. In
units $c=1$, with signature $(+,-,-,-)$ and arbitrary worldline parameters
$\theta_{1,2}$, the interaction term reads
\begin{equation}
\begin{split}
S_{\mathrm{int}}={}&-e_1e_2\int\!\dd\theta_1\!\int\!\dd\theta_2\;
x_1'^{\mu}(\theta_1)\,x'_{2\mu}(\theta_2)\\
&\times\delta\!\big[(x_1(\theta_1)-x_2(\theta_2))^2\big].
\end{split}
\label{eq:fokker_action}
\end{equation}
Its Lagrangian and Hamiltonian formulations and its symmetries have been
studied in \cite{Marnelius1974,Driver1969,GaidaTretyak1996}, and the
Lagrangian formulation of direct-interaction theories in other forms of
dynamics requires interaction terms depending on derivatives of arbitrarily
high order \cite{Duviryak1998}. For a single worldline $x^\mu(t)$ with
$x'^\mu=\dd x^\mu/\dd t$, a local action reads
\begin{equation}
S_{\mathrm{local}}=\int \dd t\,L\big(x^\mu(t),x'^\mu(t)\big),
\label{eq:local_action}
\end{equation}
whereas a nonlocal action may take the form
\begin{equation}
S_{\mathrm{nonlocal}}=\int \dd t\,\dd t'\,
K\big(x(t),x(t'),x'(t),x'(t')\big),
\label{eq:nonlocal_action}
\end{equation}
with $K$ built from Lorentz-invariant combinations such as
$(x(t)-x(t'))^2$. The Euler--Lagrange equations
$\delta S/\delta x^\mu(t)=0$ then become integro-differential, and the
evolution is no longer fixed by the instantaneous state. Nonlocality does not
by itself break Lorentz invariance or causality: this depends on the kernel, on
the domain of the interaction and on the boundary conditions. By contrast, on
noncommutative spacetime the usual Lorentz symmetry is broken, although a
twisted Poincar\'e symmetry can be retained \cite{Chaichian2004}.

\paragraph*{Symmetries and conservation laws beyond first order.}
For local first-order Lagrangians, the relation between Lorentz invariance and
conserved currents is given by Noether's theorem
\cite{Noether1918,KosmannSchwarzbach2011}. For Lagrangians of finite order
$q>1$, it extends through generalized symmetries and prolongations to the
jet space \cite{Olver1993}, and the canonical structure is that of
Ostrogradsky \cite{Ostrogradsky1850}, at the price of a Hamiltonian that is
linear in all but one of the canonical momenta and hence unbounded below
\cite{Woodard2015}. For infinitely many derivatives the
canonical formalism itself is delicate \cite{Woodard2000,BarnabyKamran2008}.
The symmetry structure, on the other hand, can be controlled order by order.
Lorentz-invariant Lagrangians that depend on higher derivatives of the
trajectory arise naturally from geometric actions involving the curvature of
the path \cite{Polyakov1986,Pisarski1986}.

\paragraph*{Why the question matters.}
In a relativistic theory the Poincar\'e charges are not optional decorations:
energy and momentum define the mass of a state, angular momentum its spin, and
the boost charge fixes the center of energy and is the only one of the ten
charges that depends explicitly on time. If a nonlocal theory is to be
physically acceptable, these charges must exist, be conserved, and reduce to
the familiar ones in the local limit. Yet for theories with infinitely many
derivatives, which are the theories motivated above, it is not even clear in
what sense a conserved current can be defined, since the canonical momenta
involve derivatives of arbitrarily high order and the solution space of the
equations of motion is itself delicate \cite{BarnabyKamran2008}.

\paragraph*{State of the art.}
Several approaches address conservation laws beyond first order. Noether's
theorem has been extended to nonlocal Lagrangians, in mechanics and in field
theory, by working directly on the space of trajectories
\cite{HerediaLlosa2021,HerediaLlosa2022,Heredia2023}, and explicit
energy--momentum and angular-momentum tensors have been constructed for
higher-derivative and nonlocal fields \cite{KrivoruchenkoTursunov2016}. For
relativistic particles, Lagrangians depending on the curvatures of the
worldline have been analyzed in the covariant (proper-time) formalism, where
mass and spin are expressed through the Frenet--Serret curvatures
\cite{Nesterenko1995,Arreaga2001}. What these works do not provide is a
treatment in which (a) the Lorentz transformation acts on the evolution
parameter itself, as it does when $t$ is the time of an inertial observer,
so that the boost generator mixes the independent and the dependent variables;
(b) the conserved boost charge is written explicitly in Ostrogradsky
variables at arbitrary order $q$; and (c) the passage $q\to\infty$ is
controlled quantitatively rather than formally.

\paragraph*{Aim and approach.}
The purpose of this paper is to fill this gap. We start from a finite-order
Lagrangian, where the machinery of prolongations applies, and regard the
nonlocal theory as the limit of an infinite-derivative description, so that
Lorentz symmetry is manifest at every order of the truncation. We work with a
single worldline, which keeps all formulas explicit; the field-theory analogue
is discussed in the conclusion.

\paragraph*{Results and novelty.}
To our knowledge, items (i)--(iii) are not available in closed form in the
literature for arbitrary $q$ in the time gauge, items (v) and (vi) are new,
and in item (vii) the jet-space realization of the algebra and the
identification of the Ostrogradsky charges with the covariant ones are new,
while the covariant expressions themselves and the integrability of the
second-order model recover known results
\cite{Nesterenko1995,Arreaga2001}.
\begin{enumerate}
\item[(i)] The prolongation of any infinitesimal Poincar\'e transformation to
arbitrary order $q$ is given in closed form (Lemmas~\ref{lem:prolongation}
and~\ref{lem:prolongation_d}).
\item[(ii)] For $L(x,\dots,x^{(q)})$ we prove an off-shell Noether identity
and obtain the explicit current $J_q$, conserved on shell when $L$ is
invariant (Theorems~\ref{thm:noether} and~\ref{thm:noether_d}).
\item[(iii)] In terms of Ostrogradsky momenta, the ten Poincar\'e charges are
given explicitly; the boost charge differs from the naive $t p-xH$ by a
term $N^k$ that we compute for every $q\ge2$ (Corollary~\ref{cor:poincare})
and identify as the boost component of an internal spin tensor. We prove that
$J_q$ is invariant under the boost it generates when the boost is prolonged to
order $2q-1$, but not under the prolongation $X^{(q)}$
(Proposition~\ref{prop:XJ}); at $q=1$, the invariance of $J_1$ singles out the
free particle (Proposition~\ref{prop:J1_invariant}).
\item[(iv)] Boost invariance forces $L$ to be independent of the positions,
and in $(1+1)$ dimensions we obtain the general solution of the invariance
condition at \emph{every} order $q$: $L=\sqrt{1-x'^2}\,F(a,\dot a,\dots,
a^{(q-2)})$, with $a$ the proper acceleration and dots denoting proper-time
derivatives (Theorem~\ref{thm:general_invariant}), an instance of the
classical theory of differential invariants \cite{Olver1995}. We explain why
the direct
integration of the invariance condition, which works for $q\le2$, becomes
impractical for $q\ge3$, and how invariant differentiation removes the
obstruction.
\item[(v)] For an infinite-derivative Lagrangian we prove that the
conservation defect of the truncated current is exactly $Q$ times the tail of
the series of equations of motion (Theorem~\ref{thm:defect}). This gives
sufficient conditions under which the limit current $J_\infty$ exists and is
exactly, not only formally, conserved.
\item[(vi)] In an explicit nonlocal model we show that on trajectories of
constant proper acceleration the series terminate, and we find exact
self-accelerated solutions with vanishing energy and momentum. Deriving the
covariant equations of motion at every order, we prove that these solutions
are linearly unstable for every nonlocal completion (Sec.~\ref{sec:model}). In the small-amplitude sector of
the same model, the convergence conditions are realized nontrivially: they
hold for the modes inside the disk of convergence of the characteristic
function and fail outside it (Sec.~\ref{sec:linear}).
\item[(vii)] In $1+1$ dimensions the boost acts on $(H,p,K)$ as the Poincar\'e
algebra, without central term (Proposition~\ref{prop:algebra}). We obtain the
momentum, the spin and the mass in closed covariant form,
$p^\mu=(a\mathcal E+\mathcal H)u^\mu-\dot{\mathcal E}n^\mu$,
$S^{01}=\mathcal E$, $M^2=(a\mathcal E+\mathcal H)^2-\dot{\mathcal E}^2$
(Theorem~\ref{thm:momentum}), derive the Zitterbewegung around the center of
energy, and recover the integrability of the second-order model
\cite{Nesterenko1995,Arreaga2001}, with an explicit phase portrait.
\end{enumerate}

The paper is organized as follows. Section~\ref{sec:boost} introduces the boost
in $(1+1)$ dimensions and its prolongation, and Sec.~\ref{sec:noether} proves
the Noether identity. Section~\ref{sec:invariance} classifies invariant
Lagrangians at every order, and Sec.~\ref{sec:nonlocal} studies the limit
$q\to\infty$ and an exactly controlled nonlocal model. Section~\ref{sec:full} extends the construction to the full Poincar\'e
group in $d$ spatial dimensions, and Sec.~\ref{sec:mass} derives the
Poincar\'e algebra, the mass and the spin in $1+1$ dimensions.
Section~\ref{sec:conclusion} collects our conclusions and discusses field
theories.

\section{Lorentz boost and its prolongation}
\label{sec:boost}

We work in $(1+1)$-dimensional Minkowski space with $c=1$. The dynamics is
described by a trajectory $x=x(t)$, where $t\equiv x^0$ is the time
coordinate of an inertial frame, used as evolution parameter. It is
\emph{not} the invariant proper time, which we denote $\sigma$,
$\dd\sigma=\sqrt{1-x'^2}\,\dd t$. We write $x^{(k)}=\dd^kx/\dd t^k$,
$x'=x^{(1)}$, and $\D$ for the total derivative with respect to $t$, acting on any function of type $F(t,x,x^{(1)},\cdots, x^{(q)})$ as:
\begin{equation}
\D F=\partial_t F+\sum_{n=0}^{q}x^{(n+1)}\partial_{x^{(n)}}F.
\end{equation}
The infinitesimal boost along $x$ is
\begin{equation}\label{eq:boost}
\bar{t}=t+\epsilon\,x,\qquad
\bar{x}=x+\epsilon\,t,
\end{equation}
with $\epsilon$ infinitesimal, generated by the vector field
\begin{equation}\label{eq:X}
X=\xi\,\partial_t+\eta\,\partial_x,\qquad \xi=x,\quad \eta=t.
\end{equation}
Since $L$ depends on derivatives of $x$, the symmetry must be prolonged to the
jet space with coordinates $(t,x,x^{(1)},\dots,x^{(q)})$:
\begin{align}\label{eq:Xq_def}
X^{(q)}&=X+\sum_{n=1}^{q}\eta^{(n)}\,\frac{\partial}{\partial x^{(n)}},\\
\eta^{(n)}&=\D\,\eta^{(n-1)}-x^{(n)}\,\D\xi,
\end{align}
with $\eta^{(0)}=\eta$. Define the \emph{characteristic} of the boost,
\begin{equation}\label{eq:Q}
Q:=\eta-\xi\,x'=t-x\,x'.
\end{equation}

\begin{lemma}\label{lem:prolongation}
The prolonged components are $\eta^{(n)}=\D^nQ+x\,x^{(n+1)}$ for all $n\ge0$.
Explicitly,
\begin{align}\label{eq:etan}
\eta^{(1)}&=1-x'^2,\\
\eta^{(n)}&=-\sum_{k=1}^{n}\binom{n}{k}\,x^{(k)}\,x^{(n+1-k)}\quad(n\ge2).
\end{align}
In particular $\eta^{(2)}=-3x'x''$, and
$\bar{x}^{(n)}=x^{(n)}+\epsilon\,\eta^{(n)}+O(\epsilon^2)$.
\end{lemma}

\begin{demonstration}
For $n=0$, $\eta^{(0)}=t=Q+x\,x'$. If the identity holds at order $n-1$,
then $\eta^{(n)}=\D\big(\D^{n-1}Q+\xi x^{(n)}\big)-x^{(n)}\D\xi
=\D^nQ+\xi\,x^{(n+1)}$. For $n=1$ this gives
$\D(t-xx')+xx''=1-x'^2$. For $n\ge2$, $\D^nt=0$ and the Leibniz rule
gives $-\D^n(xx')+x\,x^{(n+1)}=-\sum_{k=0}^{n}\binom nk
x^{(k)}x^{(n+1-k)}+x\,x^{(n+1)}$; the $k=0$ term cancels the last one.
\end{demonstration}

For $n=1$ one recovers directly
$\bar{x}'=\dd\bar{x}/\dd\bar{t}=(x'+\epsilon)/(1+\epsilon x')
=x'+\epsilon(1-x'^2)+O(\epsilon^2)$, and for $n=2$,
$\bar{x}''=x''-3\epsilon\,x'x''$. Note that $\eta^{(1)}$ contains the
inhomogeneous term $1$, absent from $\eta^{(n\ge2)}$. The prolonged field is
\begin{align}
X^{(q)}&=x\,\partial_t+t\,\partial_x+(1-x'^2)\,\partial_{x'}\nonumber\\
&\quad-\sum_{n=2}^{q}\sum_{k=1}^{n}\binom nk x^{(k)}x^{(n+1-k)}\,
\partial_{x^{(n)}} .
\label{eq:Xq}
\end{align}

\section{Noether identity and the boost current $J_q$}
\label{sec:noether}

Let $L=L(x,x^{(1)},\dots,x^{(q)})$, $q\ge1$, with no explicit dependence on
$t$, and write $L_n=\partial L/\partial x^{(n)}$ ($L_0=\partial L/\partial
x$). The Euler--Lagrange expression is
\begin{equation}\label{eq:EL}
E(L)=\sum_{n=0}^{q}(-1)^n\,\D^n L_n .
\end{equation}

\begin{theorem}[Boost Noether identity]\label{thm:noether}
Define
\begin{equation}\label{eq:Jq_compact}
J_q=x\,L+\sum_{n=1}^{q}\sum_{j=0}^{n-1}(-1)^j\,
\D^{\,n-1-j}(Q)\;\D^{\,j}\!\big(L_n\big).
\end{equation}
Then, for every smooth trajectory $x(t)$,
\begin{equation}\label{eq:identity}
\D J_q = X^{(q)}L+x'\,L-Q\,E(L).
\end{equation}
\end{theorem}

Relabeling $m=n-1-j$, the current takes the form
\begin{align}\label{eq:Jq_ostro}
J_q&=x\,L+\sum_{m=0}^{q-1}\big(\D^mQ\big)\,P^{(m)},\\
P^{(m)}&=\sum_{n=m+1}^{q}(-1)^{n-m-1}\D^{\,n-m-1}L_n,
\end{align}
where $P^{(m)}$ is the Ostrogradsky momentum conjugate to $x^{(m)}$
\cite{Ostrogradsky1850,Woodard2015}. Expanding $\D^m Q$ with
$\D Q=1-\D(xx')$ and $\D^mQ=-\D^m(xx')$ for $m\ge2$ gives the explicit form
\begin{align}
J_q&=xL+Q\,L_1\nonumber\\
&\quad+\sum_{n=2}^{q}\Bigg[(-1)^{n-1}Q\,\D^{n-1}L_n\nonumber\\
&\qquad+\sum_{k=1}^{n-1}(-1)^{k}\,\D^{n-k}(xx')\,\D^{k-1}L_n\nonumber\\
&\qquad+(-1)^{n}\D^{n-2}L_n\Bigg],
\label{eq:Jq}
\end{align}
and for $q=1$ it reduces to
\begin{equation}\label{eq:J1}
J_1=x\,L+(t-x\,x')\,\frac{\partial L}{\partial x'} .
\end{equation}

\begin{demonstration}
\emph{Step 1.} Since $\partial_t L=0$, $\xi\,\D L=\sum_n x\,x^{(n+1)}L_n$,
so Lemma~\ref{lem:prolongation} gives
\begin{equation}
X^{(q)}L=\sum_{n=0}^{q}\eta^{(n)}L_n
=\sum_{n=0}^{q}\big(\D^nQ\big)L_n+x\,\D L,
\end{equation}
and therefore $X^{(q)}L+x'L=\sum_{n=0}^q(\D^nQ)L_n+\D(xL)$.

\emph{Step 2.} For smooth functions $f,g$ one has, by telescoping,
\begin{equation}\label{eq:ibp}
(\D^nf)\,g=(-1)^n f\,\D^ng
+\D\sum_{j=0}^{n-1}(-1)^j\,\D^{n-1-j}f\;\D^jg .
\end{equation}
Applying this with $f=Q$, $g=L_n$ and summing over $n$ yields
$\sum_n(\D^nQ)L_n=Q\,E(L)+\D\sum_{n\ge1}\sum_{j}(-1)^j\D^{n-1-j}Q\,\D^jL_n$.

\emph{Step 3.} Combining Steps 1 and 2, $X^{(q)}L+x'L=Q\,E(L)+\D J_q$,
which is \eqref{eq:identity}. The explicit form \eqref{eq:Jq} follows from the
expression of $\D^mQ$ given above.
\end{demonstration}

\begin{corollary}[Conservation]\label{cor:conservation}
If $L$ is boost invariant in the sense
\begin{equation}\label{eq:invariance}
X^{(q)}L=-x'\,L ,
\end{equation}
then $\D J_q=-Q\,E(L)$, so $J_q$ is conserved on every solution of
$E(L)=0$. Conversely, if $J_q$ is conserved on all solutions of $E(L)=0$,
then $X^{(q)}L+x'L$ vanishes on the solutions, i.e.\ \eqref{eq:invariance}
holds modulo the equations of motion.
\end{corollary}

\begin{remark}
Condition \eqref{eq:invariance} is the requirement that
$L\,\dd t$ be invariant, since $\dd\bar t=(1+\epsilon x')\dd t$:
$\delta S=\epsilon\int\dd t\,(X^{(q)}L+x'L)$. On shell, \eqref{eq:identity}
reads $\D J_q=X^{(q)}L+x'L$, the infinitesimal counterpart of the conservation
law. A second, direct proof of Corollary~\ref{cor:conservation}, obtained by
differentiating \eqref{eq:Jq} term by term, is given in
Appendix~\ref{app:direct}. 
\end{remark}

\section{Boost-invariant Lagrangians}
\label{sec:invariance}

\begin{corollary}\label{cor:no_x}
If \eqref{eq:invariance} holds identically on the jet space, then $L$ does not
depend on $x$, and \eqref{eq:invariance} reduces to
\begin{equation}\label{eq:invariance_red}
\sum_{n=1}^{q}\eta^{(n)}\,\frac{\partial L}{\partial x^{(n)}}+x'L=0 .
\end{equation}
\end{corollary}

\begin{demonstration}
In $X^{(q)}L+x'L=t\,L_0+\sum_{n\ge1}\eta^{(n)}L_n+x'L$ the variable $t$
appears explicitly only in the first term, because $L$ and the $\eta^{(n)}$
do not depend on $t$. The condition must hold for all $t$, hence
$L_0=\partial L/\partial x=0$.
\end{demonstration}

\begin{example}[$q=1$: free particle]
Equation \eqref{eq:invariance_red} reads $(1-x'^2)L'+x'L=0$, whose solution
is $L=-m\sqrt{1-x'^2}$, with $m$ an integration constant. Then \eqref{eq:J1}
gives
\begin{equation}
J_1=\frac{m\,(t x'-x)}{\sqrt{1-x'^2}}=t\,p-x\,H,
\end{equation}
with $p=m\gamma x'$ and $H=m\gamma$, which is conserved since $x''=0$ on
shell. Thus, for a first-order Lagrangian, strict boost invariance leaves no
freedom beyond the free particle.
\end{example}

\paragraph*{Partial conclusion for $q=1$: when is $J_1$ boost invariant?}
The previous example starts from an invariant Lagrangian. One may ask the
converse question: for which first-order Lagrangians $L(x,x')$ is the
function $J_1=xL+(t-xx')\,\partial L/\partial x'$ itself invariant under
the boost, $X^{(1)}J_1=0$, with $X^{(1)}=x\,\partial_t+t\,\partial_x
+(1-x'^2)\,\partial_{x'}$?

\begin{proposition}\label{prop:J1_invariant}
Let $L=L(x,x')$ be smooth for $x\neq0$, $|x'|<1$. Then $X^{(1)}J_1=0$
identically if and only if $L$ satisfies the system
\begin{align}
\frac{\partial^2L}{\partial x\,\partial x'}&=0, \label{eq:J1inv_a}\\
x'(1-x'^2)\,\frac{\partial^2L}{\partial x'^2}-\frac{\partial L}{\partial x'}
&=0, \label{eq:J1inv_b}\\
L+x\,\frac{\partial L}{\partial x}-x'\,\frac{\partial L}{\partial x'}
+(1-x'^2)\,\frac{\partial^2L}{\partial x'^2}&=0, \label{eq:J1inv_c}
\end{align}
whose general solution is
\begin{equation}\label{eq:J1inv_sol}
L=-m\sqrt{1-x'^2}+\frac{\kappa}{x},\qquad m,\kappa\in\mathbb R .
\end{equation}
\end{proposition}

\begin{demonstration}
Write $v=x'$ and $L_v=\partial L/\partial v$. Since $L$ does not depend on
$t$, $\partial_t J_1=L_v$,
$\partial_xJ_1=L+xL_x-vL_v+(t-xv)L_{xv}$ and
$\partial_vJ_1=(t-xv)L_{vv}$. Hence
\begin{equation}
\begin{split}
X^{(1)}J_1={}&t^2L_{xv}+x\big[L_v-v(1-v^2)L_{vv}\big]\\
&+t\big[L+xL_x-vL_v\\
&\qquad-xvL_{xv}+(1-v^2)L_{vv}\big].
\end{split}
\end{equation}
This polynomial in $t$ vanishes identically iff its three coefficients
vanish, which gives \eqref{eq:J1inv_a}--\eqref{eq:J1inv_c}. By
\eqref{eq:J1inv_a}, $L=A(x)+B(v)$. Equation \eqref{eq:J1inv_b},
$B''/B'=1/[v(1-v^2)]$, integrates to $B'=m\,v/\sqrt{1-v^2}$, i.e.\
$B=-m\sqrt{1-v^2}+D$. Then $B-vB'+(1-v^2)B''=D$, and \eqref{eq:J1inv_c}
reduces to $(xA)'=-D$, so $A=-D+\kappa/x$, which gives
\eqref{eq:J1inv_sol}.
\end{demonstration}

The singular point $x=0$ is the position of the origin of the boost, which
is fixed by the choice of generator \eqref{eq:boost}; the result is therefore
specific to the boost about the origin, and a boost about another point $x_0$
gives $\kappa/(x-x_0)$. The term $\kappa/x$ is trivial: since $J_1$ is linear in $L$ and
$\partial(\kappa/x)/\partial x'=0$, it contributes the constant $\kappa$ to
$J_1$, which is invariant but carries no dynamics. Up to this term, the
invariance of the current under the boost singles out exactly the free
relativistic particle, the same Lagrangian as the invariance of
$L\,\dd t$. Note also that \eqref{eq:J1inv_b} is a consequence of the
invariance condition $(1-v^2)L_v+vL=0$: differentiating it gives
$(1-v^2)L_{vv}=vL_v-L$, and eliminating $L$ gives \eqref{eq:J1inv_b}. At first order,
therefore, three properties coincide modulo trivial terms: the invariance of
$L\,\dd t$, the conservation of $J_1$, and the invariance of $J_1$ under the
boost. Proposition~\ref{prop:XJ} below shows that the last
equivalence fails for $q\ge2$ if the boost is prolonged only to order $q$.

\begin{example}[$q=2$: proper acceleration]
Let $s=\sqrt{1-x'^2}$ and $a=x''/s^3$ be the proper acceleration. From
\eqref{eq:Xq}, $X^{(2)}s=-x's$ and $X^{(2)}a=0$, hence
\begin{equation}
L=\sqrt{1-x'^2}\;f(a)
\end{equation}
satisfies $X^{(2)}L=f\,X^{(2)}s=-x'L$ for any function $f$, i.e.\
\eqref{eq:invariance}. This is the statement that
$S=\int f(a)\,\dd\sigma$ is built from Lorentz scalars. Conversely, for $q=2$
Eq.~\eqref{eq:invariance_red} is a first-order linear PDE in the two variables
$(x',x'')$; $s$ is a particular solution and $a$ is a first integral of its
characteristics, so $L=s\,f(a)$ is its general solution. The corresponding
conserved current is \eqref{eq:Jq} with $q=2$.
\end{example}

\paragraph*{What blocks a direct treatment for $q\ge3$.}
Writing $L=s\,G$ with $G=G(x',\dots,x^{(q)})$, the invariance condition
\eqref{eq:invariance_red} becomes, since $X^{(q)}s=-x's$,
\begin{equation}\label{eq:VG}
V G=0,\qquad V:=\sum_{n=1}^{q}\eta^{(n)}\frac{\partial}{\partial x^{(n)}} .
\end{equation}
For $q=1,2$ the characteristic system $\dd x^{(n)}/\dd\lambda=\eta^{(n)}$ can
be integrated by hand. For $q\ge3$ it becomes a system of $q$ coupled
nonlinear equations: by Lemma~\ref{lem:prolongation} each $\eta^{(n)}$ is
quadratic and couples all orders $x^{(k)}$, $k\le n$, and the number of first
integrals to be found grows with $q$. Integrating the characteristics
directly is then impractical. The obstruction is technical, not structural: the
first integrals can be generated recursively, without solving the
characteristic system, by \emph{invariant differentiation} along the proper
time.

\begin{lemma}[Invariant differentiation]\label{lem:invdiff}
For every function $F$ on the jet space,
$X^{(r+1)}(\D F)=\D(X^{(r)}F)-(\D\xi)\,\D F$, with $\D\xi=x'$ for the boost.
Consequently, if $I(x',\dots,x^{(r)})$ satisfies $X^{(r)}I=0$, then
$D_\sigma I:=s^{-1}\D I$ satisfies $X^{(r+1)}D_\sigma I=0$.
\end{lemma}

\begin{demonstration}
Write $F_n=\partial F/\partial x^{(n)}$. Using $[\partial_{x^{(n)}},\D]=
\partial_{x^{(n-1)}}$ ($n\ge1$), $[\partial_t,\D]=0$ and
\eqref{eq:Xq_def}, one finds $X^{(r+1)}\D F-\D X^{(r)}F=\sum_{n\ge0}
(\eta^{(n+1)}-\D\eta^{(n)})F_n-(\D\xi)\partial_t F=-(\D\xi)\D F$. Then
$X^{(r+1)}(s^{-1}\D I)=-s^{-2}(X^{(1)}s)\,\D I+s^{-1}\big(\D X^{(r)}I-x'\D I
\big)=s^{-1}x'\D I-s^{-1}x'\D I=0$.
\end{demonstration}

\begin{theorem}[Invariant Lagrangians, $1+1$]\label{thm:general_invariant}
Let $a_0:=a=x''/s^3$ and $a_k:=D_\sigma^k a$. A Lagrangian
$L(x,x',\dots,x^{(q)})$, $q\ge2$, satisfies \eqref{eq:invariance}
identically on the domain $|x'|<1$ if and only if
\begin{equation}\label{eq:general_invariant}
L=\sqrt{1-x'^2}\;F\big(a,a_1,\dots,a_{q-2}\big)
\end{equation}
for some function $F$ of $q-1$ variables. For $q=1$, $L=-m\sqrt{1-x'^2}$.
\end{theorem}

\begin{demonstration}
By Corollary~\ref{cor:no_x}, $L$ does not depend on $x$, and the condition is
\eqref{eq:VG}. By the $q=2$ example and Lemma~\ref{lem:invdiff}, $Va_k=0$ for
$0\le k\le q-2$. By induction, $a_k=s^{-(k+3)}x^{(k+2)}+g_k(x',\dots,
x^{(k+1)})$, so the map $(x',x'',\dots,x^{(q)})\mapsto(x',a_0,\dots,a_{q-2})$
is a global diffeomorphism of $\{|x'|<1\}\times\mathbb R^{q-1}$ (it is
triangular with nonvanishing diagonal). In the new coordinates $V$ annihilates
every $a_k$, and $Vx'=\eta^{(1)}=s^2$, so $V=s^2\,\partial/\partial x'$.
Hence $VG=0$ if and only if $G$ is independent of $x'$, i.e.\ a function of
$(a_0,\dots,a_{q-2})$ only.
\end{demonstration}

Thus the only Lorentz-invariant data of a worldline in $1+1$ dimensions are
the proper acceleration and its proper-time derivatives, and every invariant
Lagrangian of order $q$ is equivalent to the action $\int\dd\sigma\,F(a,\dot
a,\dots)$, with $\dot{}=\dd/\dd\sigma$. Theorem~\ref{thm:general_invariant}
is the $(1+1)$-dimensional instance of the classical result that the
differential invariants of a transformation group acting on curves are
generated by a curvature and its derivatives with respect to an invariant
arclength \cite{Olver1995}. We have included the short self-contained proof
because it answers concretely why the direct method stops at $q=2$, and
because Lemma~\ref{lem:invdiff} is used again below. In $d$ spatial
dimensions the role of $a$ is played by the $d$ Frenet--Serret curvatures of
the worldline and their derivatives \cite{Nesterenko1995,Arreaga2001}.
Lagrangians of this type are one-dimensional analogues of the
curvature-dependent actions of Refs.~\cite{Polyakov1986,Pisarski1986}.

\paragraph*{Invariance of the current requires the full Ostrogradsky state.}
Conservation of $J_q$ and invariance of $J_q$ under the boost it generates
are distinct properties. For $q=1$ they coincide
(Proposition~\ref{prop:J1_invariant}). For $q\ge2$, however, the
Ostrogradsky momenta bring derivatives of $L_n$ into $J_q$, which therefore
depends on $x^{(2q-1)}$, beyond the order $q$ of the Lagrangian. The state of
a system of order $q$ is a point of the $2q$-dimensional Ostrogradsky phase
space, with coordinates $(x,\dots,x^{(2q-1)})$, so the boost has to be
represented on functions of $J_q$'s arguments by $X^{(2q-1)}$, not by
$X^{(q)}$.

\begin{proposition}[Invariance of $J_q$]\label{prop:XJ}
If $L$ satisfies \eqref{eq:invariance} identically, then
\begin{equation}\label{eq:XJzero}
X^{(2q-1)}J_q=0
\end{equation}
identically on the jet space. Consequently
\begin{equation}\label{eq:XqJ}
X^{(q)}J_q=-\sum_{n=q+1}^{2q-1}\eta^{(n)}\,
\frac{\partial J_q}{\partial x^{(n)}} ,
\end{equation}
which in general does not vanish for $q\ge2$.
\end{proposition}

\begin{demonstration}
Write $X^{(r)}=\mathrm{pr}\,v_Q+\xi\,\D$, where $\mathrm{pr}\,v_Q=\sum_n
(\D^nQ)\,\partial_{x^{(n)}}$ is the prolonged evolutionary field with
characteristic $Q$ \cite{Olver1993}; this follows from
Lemma~\ref{lem:prolongation}.

\emph{Step 1: $X^{(2q)}E(L)=0$.} Invariance means $X^{(q)}L+(\D\xi)L=0$, i.e.\
$\mathrm{pr}\,v_Q(L)=-\D(\xi L)$, a total derivative, so that
$E(\mathrm{pr}\,v_QL)=0$. On the other hand, integrating by parts as in Step~2
of the proof of Theorem~\ref{thm:noether}, $\mathrm{pr}\,v_QL=\sum_n(\D^nQ)L_n
=Q\,E(L)+\D(\cdots)$, hence $E(\mathrm{pr}\,v_QL)=E\big(Q\,E(L)\big)
=D_Q^*E(L)+D_{E}^*(Q)$ by the product rule for the Euler operator, where $D_P$
denotes the Fr\'echet derivative of $P$ and $D_P^*$ its adjoint
\cite[Ch.~5]{Olver1993}. Since $E(L)$ is an Euler--Lagrange expression, $D_E$
is self-adjoint (Helmholtz conditions, \cite[Ch.~5]{Olver1993}), and
$D_E(Q)=\mathrm{pr}\,v_Q\,E(L)$. Therefore
$\mathrm{pr}\,v_Q\,E(L)=-D_Q^*E(L)$. For $Q=t-xx'$, $D_Q(w)=-x'w-xw'$ and
$D_Q^*(E)=-x'E+\D(xE)=x\,\D E$. Hence $X^{(2q)}E=-x\,\D E+\xi\,\D E=0$.

\emph{Step 2: $X^{(2q-1)}J_q$ is constant.} By Lemma~\ref{lem:invdiff},
$\D(XF)=X(\D F)+(\D\xi)\,\D F$. With $\D J_q=-Q\,E$ (Theorem~\ref{thm:noether}
and invariance) and $XQ=-x'Q$ (direct computation),
$\D(XJ_q)=-X(QE)-x'QE=x'QE-Q\,XE-x'QE=0$ identically. A function $K$ on a
finite-order jet space with $\D K\equiv0$ is constant: the highest derivative
$x^{(r+1)}$ appears in $\D K$ only through $x^{(r+1)}\partial K/\partial
x^{(r)}$, so $\partial K/\partial x^{(r)}=0$, and by descending induction $K$
depends on no variable. The domain $|x'|<1$ is connected, so
$X^{(2q-1)}J_q\equiv c$.

\emph{Step 3: $c=0$.} Evaluate at the jet of the hyperbola
$x(t)=\sqrt{R^2+t^2}$, an orbit of the boost. Along it $Q\equiv0$, so
all $\D^nQ$ vanish there, $\mathrm{pr}\,v_QJ_q=0$, and
$XJ_q=\xi\,\D J_q=-\xi\,Q\,E=0$. Hence $c=0$.
\end{demonstration}

For $q=1$, Eq.~\eqref{eq:XqJ} has an empty sum, so $X^{(1)}J_1=0$; for the
free particle this is checked directly from $X^{(1)}(t x'-x)=-x'(t
x'-x)$ and $X^{(1)}(1/s)=x'/s$. For $q=2$ and $L=s\,f(a)$, $J_2$ depends on
$x'''$ only through $\dot a$ in $P^{(0)}$, with $\partial J_2/\partial
x'''=-Q\,f''(a)/s^5$, so
\begin{equation}\label{eq:XJ2}
X^{(2)}J_2=-\frac{(4x'x'''+3x''^2)\,Q\,f''(a)}{(1-x'^2)^{5/2}},
\end{equation}
which is nonzero whenever $f''\neq0$ and $Q\neq0$. A test of invariance at
order $q$ thus discards part of the state, and conservation of $J_q$ does not
imply its invariance under $X^{(q)}$. In the limit $q\to\infty$ of
Sec.~\ref{sec:nonlocal}, the invariance of $J_\infty$ requires the
prolongation of infinite order, consistent with the nonlocal character of the
limiting theory.  The proof uses the dimension only
through the explicit forms of $Q$ and $D_Q^*$. For a boost along $x^k$ in $d$
dimensions, the orbit $x^k=\sqrt{R^2+t^2}$, $x^j=\mathrm{const}$ ($j\ne k$),
again has $Q^i\equiv0$, so Step~3 carries over; Steps~1 and~2 require the
analogous computation with $E_i$ transforming as a covector, which we have
not carried out. Propositions~\ref{prop:XJ} and~\ref{prop:algebra} are
therefore established here in $1+1$ dimensions only.

\section{The infinite-derivative limit}
\label{sec:nonlocal}

When the kernel of a nonlocal action such as \eqref{eq:nonlocal_action} is
analytic, it can be expanded formally as a series of local terms of increasing
derivative order. For a worldline, the natural Lorentz-invariant building
blocks are the boost-invariant Lagrangians of Sec.~\ref{sec:invariance},
constructed from $\sigma$-derivatives. The key observation is that $J_q$ in
\eqref{eq:Jq_compact} is \emph{linear} in $L$.

\begin{proposition}\label{prop:partial}
Let $L^{[q]}=\sum_{n=1}^{q}\ell_n$, where each $\ell_n$ is boost invariant,
$X^{(n)}\ell_n=-x'\ell_n$, and of order at most $n$. Then
\begin{equation}
J_q\big[L^{[q]}\big]=\sum_{n=1}^{q}J_n[\ell_n],\qquad
\D J_q\big[L^{[q]}\big]=-Q\,E\big(L^{[q]}\big).
\end{equation}
\end{proposition}

\begin{demonstration}
Since $\ell_n$ does not depend on $x^{(m)}$ for $m>n$, $X^{(q)}\ell_n=
X^{(n)}\ell_n$, so $L^{[q]}$ satisfies \eqref{eq:invariance} by linearity and
Corollary~\ref{cor:conservation} applies. Moreover $J_q$ is linear in $L$, and
the terms of $J_q[\ell_n]$ involving $\partial\ell_n/\partial x^{(m)}$ with
$m>n$ vanish, so $J_q[\ell_n]=J_n[\ell_n]$.
\end{demonstration}

Thus $J_q[L^{[q]}]$ is the $q$-th partial sum of a series, and the Noether
current of the infinite-derivative Lagrangian $L^{[\infty]}=\sum_n\ell_n$ is
\emph{defined} as its limit when it exists, $J_\infty=\sum_{n\ge1}J_n[\ell_n]$.
By Theorem~\ref{thm:general_invariant}, in $1+1$ dimensions every term has the
form $\ell_n=s\,F_n(a,a_1,\dots,a_{n-2})$, so $L^{[\infty]}\dd t=
\sum_nF_n\,\dd\sigma$ is a series of local Lorentz scalars on the worldline.

\subsection{The limit $q\to\infty$: conservation defect of the truncations}

A naive argument would state that $\D J_\infty=-Q\,E(L^{[\infty]})$ follows
from Proposition~\ref{prop:partial} by passing to the limit. This is not
correct as it stands, for a simple reason: a solution of the infinite-order
equation is \emph{not} a solution of any truncated equation, so the truncated
currents are never conserved on it. The following identity makes this
precise.

\begin{theorem}[Conservation defect]\label{thm:defect}
Let $x(t)$ be a smooth trajectory on an interval $I$ such that the series
$\sum_{n\ge1}E(\ell_n)$ converges pointwise on $I$ to
$E_\infty:=E(L^{[\infty]})$, and define the tail
$R_q:=\sum_{n>q}E(\ell_n)=E_\infty-E(L^{[q]})$. Then, for every $q$,
\begin{equation}\label{eq:defect}
\D J_q\big[L^{[q]}\big]=Q\,R_q-Q\,E_\infty .
\end{equation}
In particular, on a solution of the infinite-order equation, $E_\infty=0$,
\begin{equation}\label{eq:drift}
J_q\big[L^{[q]}\big](t_2)-J_q\big[L^{[q]}\big](t_1)
=\int_{t_1}^{t_2}\!Q\,R_q\,\dd t .
\end{equation}
\end{theorem}

\begin{demonstration}
By Proposition~\ref{prop:partial}, $\D J_q[L^{[q]}]=-Q\,E(L^{[q]})=
-Q(E_\infty-R_q)$. Integrating gives \eqref{eq:drift}.
\end{demonstration}

The truncated charge therefore drifts, and its drift is controlled
exactly by the tail of the equations of motion, weighted by the characteristic
$Q=t-xx'$. Exact conservation in the limit follows under explicit, checkable
conditions.

\begin{corollary}[Exact conservation of $J_\infty$]\label{cor:Jinfty}
Let $x(t)$ solve $E_\infty=0$ on $I$ in the sense of
Theorem~\ref{thm:defect}, and assume that (i) the series
$J_\infty=\sum_nJ_n[\ell_n]$ converges at every point of $I$, and (ii)
$\int_{t_1}^{t_2}|Q\,R_q|\,\dd t\to0$ as $q\to\infty$ for all
$t_1,t_2\in I$ (for instance, $|R_q|\le g$ with $g$ locally integrable,
by dominated convergence). Then $J_\infty$ is constant on $I$.
\end{corollary}

\begin{demonstration}
Take $q\to\infty$ in \eqref{eq:drift}: the left-hand side tends to
$J_\infty(t_2)-J_\infty(t_1)$ by (i), the right-hand side to $0$ by
(ii).
\end{demonstration}

Theorem~\ref{thm:defect} uses only the linearity of $J$ in $L$ and the
Noether identity. It therefore holds verbatim for any Poincar\'e generator of
Theorem~\ref{thm:noether_d} \emph{under which every term $\ell_n$ is
invariant}, with $Q$ replaced by the corresponding characteristic $Q^i$; for
time translations, which leave every autonomous $\ell_n$ invariant, $Q=-x'$
and $J=-H$. Two remarks
clarify the scope of this result. First, no convergence of the
series of derivatives $\sum_n\D J_n[\ell_n]$ is required: condition (ii) is a
condition on the equations of motion, which is the natural object in the
analysis of infinite-order equations \cite{BarnabyKamran2008}. Second,
conditions (i)--(ii) constrain the pair (Lagrangian, trajectory) and not the
Lagrangian alone. Since $\ell_n$ has order $n$, $P^{(0)}[\ell_n]$ contains
$\D^{n-1}\partial\ell_n/\partial x^{(n)}$, so $J_n[\ell_n]$ involves derivatives
of the trajectory up to order $2n-1$. Heuristically, for a trajectory that is
merely real-analytic, Cauchy estimates allow $|x^{(k)}|$ to grow like
$k!\,\rho^{-k}$, so that one expects convergence to require coefficients
decaying faster than this factorial growth, whereas for trajectories of
exponential type, $|a_k|\le C M^k$, geometric decay should suffice. We do not
attempt to make these estimates rigorous in general; Sec.~\ref{sec:linear}
shows how they become sharp in an explicit case. This is the worldline
counterpart of the statement that infinite-derivative equations are well posed
only on restricted function classes
\cite{BarnabyKamran2008,CarlssonPradoReyes2016,Heredia2023}. Finally, the
theorem clarifies the status of the additional, ghost-like solutions that
every truncation carries \cite{Ostrogradsky1850,Woodard2000,Woodard2015}:
these solve $E(L^{[q]})=0$, so $J_q$ is conserved on them, but in general
$R_q\ne0$ there, and they need not survive the limit $q\to\infty$.

\subsection{An exactly controlled model}\label{sec:model}

Consider the nonlocal worldline action
\begin{equation}\label{eq:model}
S=\int\dd\sigma\Big[-m+\frac12\sum_{k\ge0}c_k\,a_k^2\Big],
\end{equation}
the worldline analogue of $\phi\,F(\Box)\phi$, i.e.\ $\ell_1=-ms$ and
$\ell_{k+2}=\tfrac12c_k\,s\,a_k^2$, of order $k+2$. We look for solutions with
constant proper acceleration, $a_k\equiv0$ for $k\ge1$; in $1+1$ dimensions
these are straight lines ($a=0$) and hyperbolas.

\emph{Termination of the series.} For $k\ge1$, $\partial\ell_{k+2}/\partial
x^{(n)}=c_k\big(\tfrac12a_k^2\,\partial s/\partial x^{(n)}+s\,a_k\,\partial
a_k/\partial x^{(n)}\big)$ contains the factor $a_k$, which vanishes
identically along the trajectory together with all its $t$-derivatives.
Hence $E(\ell_{k+2})=0$ and $J_{k+2}[\ell_{k+2}]=0$ on such trajectories: the
tails $R_q$ vanish for $q\ge2$, conditions (i)--(ii) of
Corollary~\ref{cor:Jinfty} hold trivially for \emph{any} coefficients $c_k$,
and all charges are those of the $q=2$ Lagrangian $s\,f(a)$ with
$f(a)=-m+\tfrac12c_0a^2$.

\emph{Charges of $s\,f(a)$.} For $L=s f(a)$, with $\gamma=1/s$ and
$v=x'$, the Ostrogradsky momenta are $P^{(1)}=f'(a)/s^2$ and
$P^{(0)}=\gamma\big[v(af'-f)-f''\,\dot a\big]$, and the Hamiltonian is
\begin{equation}\label{eq:H_sf}
H=\gamma\big[a f'(a)-f(a)\big]-\gamma\,v\,f''(a)\,\dot a .
\end{equation}
For $f=-m$ this gives $H=m\gamma$. The time-translation counterpart of
Theorem~\ref{thm:noether} (the identity $\D H=-x'E$, valid off shell for
every autonomous Lagrangian; see Corollary~\ref{cor:poincare} for its
derivation in $d$ dimensions) gives $E=-\D H/x'$. On a trajectory with
constant $a$, using $\D\gamma=va$, we obtain
\begin{equation}\label{eq:EOM_const}
E=-a\big[a f'(a)-f(a)\big].
\end{equation}

\emph{Exact solutions.} For the model \eqref{eq:model},
$af'-f=m+\tfrac12c_0a^2$. Straight lines are always solutions, with
$H=m\gamma$. Hyperbolic motion is an exact solution of the full nonlocal
theory if and only if
\begin{equation}
c_0<0,\qquad a^2=\frac{2m}{|c_0|},
\end{equation}
independently of $c_k$ for $k\ge1$. On these self-accelerated solutions
$H=0$ and $p=P^{(0)}=0$, while the boost charge \eqref{eq:K} reduces to
$K=N=(1-v^2)P^{(1)}=f'(a)=c_0a$, a nonzero constant, as required by
Corollary~\ref{cor:conservation}. (Note that $Q=t-xx'$ vanishes on a
hyperbola centered at the origin, which is an orbit of the boost.) In the
language of Eq.~\eqref{eq:spin}, the whole Poincar\'e charge of these
solutions is internal spin, $M^{01}=S^{01}=c_0a$. Relativistic particles
whose action contains, besides the length, a term quadratic in the curvature
(``rigid particles'') have been studied extensively
\cite{Nesterenko1989,Plyushchay1988,Plyushchay1989,Nesterenko1995}; the
point here is that the self-accelerated solutions of the truncated model
survive the nonlocal completion, for any $c_k$, rather than being artifacts
of the truncation.

\emph{Covariant equations of motion.} Stability is most easily analyzed in
covariant form. For an invariant Lagrangian $L=s\,F(a,a_1,\dots,a_r)$, define
\begin{align}
\pi_j&=\sum_{k=j+1}^{r}\Big(-\frac{\dd}{\dd\sigma}\Big)^{k-j-1}F_{a_k},
\label{eq:pi_H}\\
\mathcal H&=\sum_{j=0}^{r-1}\pi_j\,a_{j+1}-F ,\nonumber
\end{align}
the Ostrogradsky momenta and Hamiltonian of $F$ regarded as a
higher-derivative Lagrangian for the ``coordinate'' $a(\sigma)$, and its
variational derivative
$\mathcal E=\sum_{k=0}^{r}(-\dd/\dd\sigma)^kF_{a_k}$.

\begin{proposition}[Covariant form]\label{prop:cov_eom}
For every order $r$,
\begin{equation}\label{eq:cov_eom}
E(L)=\ddot{\mathcal E}-a^2\mathcal E-a\,\mathcal H .
\end{equation}
\end{proposition}

The proof, by a normal variation of the worldline, is given in
Appendix~\ref{app:cov}. Equation \eqref{eq:cov_eom} is the Minkowskian
analogue of the equation of planar elastic curves, with the sign of the $a^2$
term reversed (cf.\ the relativistic elastica of Ref.~\cite{Dereli1990}). For
$r=0$, $\mathcal E=F'(a)$ and $\mathcal H=-F$, and \eqref{eq:cov_eom} reduces
to \eqref{eq:EOM_const} for constant $a$. 

\emph{Linear instability of the self-accelerated solutions.} Perturb the
self-accelerated solution of the model \eqref{eq:model},
$a=a_*+\delta(\sigma)$, $a_*^2=2m/|c_0|$. Then $a_k=\delta^{(k)}$ for $k\ge1$,
so $\mathcal E=c_0a_*+f(-\dd^2/\dd\sigma^2)\,\delta$ with
$f(z)=\sum_kc_kz^k$, while every term of $\sum_j\pi_ja_{j+1}$ is quadratic in
$\delta$, so $\mathcal H=-F+O(\delta^2)$. Inserting into \eqref{eq:cov_eom},
the terms of order zero cancel because $F(a_*)=c_0a_*^2=-2m$, the terms of
order $\delta$ without derivatives cancel for the same reason, and the
linearized equation is
\begin{equation}\label{eq:lin_stab}
\Big(\frac{\dd^2}{\dd\sigma^2}-a_*^2\Big)\,
f\Big(-\frac{\dd^2}{\dd\sigma^2}\Big)\,\delta=0 .
\end{equation}
For every choice of the coefficients $c_k$ (provided $f(-a_*^2)$ converges,
which is automatic for entire $f$), the first factor yields the modes
$\delta\propto e^{\pm a_*\sigma}$: the self-accelerated solution is linearly
unstable, with growth rate equal to its own proper acceleration, for every
nonlocal completion. For $c_k=0$, $k\ge1$, this agrees with the exact first
integral of $c_0\ddot a=ma+\tfrac12c_0a^3$, for which $a_*$ is a saddle
point. The additional modes, given by the zeros of $f(-\lambda^2)$, are those
of the nonlocal completion.

\subsection{A nontrivial test of the limit: the small-amplitude sector}
\label{sec:linear}

On constant-acceleration trajectories the series of Sec.~\ref{sec:model}
terminate, so they do not test the convergence conditions of
Corollary~\ref{cor:Jinfty}. A nontrivial test is provided by the
small-amplitude sector of the model. To quadratic order in $x',x'',\dots$ we
have $s=1-\tfrac12x'^2+\dots$ and $a_k=x^{(k+2)}+\dots$, so that, dropping the
constant $-m$,
\begin{equation}\label{eq:quad}
L^{[\infty]}_{(2)}=\frac12\sum_{n\ge1}b_n\big(x^{(n)}\big)^2,\qquad
b_1=m,\quad b_{k+2}=c_k .
\end{equation}
We write $\tilde\ell_n=\tfrac12b_n(x^{(n)})^2$ for these terms, which are
invariant under time translations but not under boosts: in this sector the
Lorentz boost reduces to a Galilean boost, and the test concerns the energy,
for which Theorem~\ref{thm:defect} holds with $Q=-x'$. Two limitations should
be stated explicitly. First, the limit $q\to\infty$ of the \emph{boost}
charge is tested nontrivially nowhere in this paper; Sec.~\ref{sec:model}
tests it only on trajectories where the series terminate. Second, the
exponential modes below are exact solutions of the quadratic theory
\eqref{eq:quad} but only approximate solutions of the full model;
heuristically, the nonlinear terms, of relative order $|A|^2|\lambda|^2$,
shift the characteristic roots $z$ by a correspondingly small amount, so that
the
criterion $|z|<\rho$ derived below is valid for modes whose root lies at a
finite distance from the circle $|z|=\rho$. Define $g(z)=\sum_{n\ge1}b_nz^n=
mz+z^2f(z)$. Since $E(\tilde\ell_n)=(-1)^nb_nx^{(2n)}$, the exponential $e^{\lambda t}$ gives $E=g(z)\,x$ with
$z=-\lambda^2$: the truncated and infinite-order equations have
characteristic functions $g^{[q]}(z)=\sum_{n\le q}b_nz^n$ and $g(z)$, and
the tail of the equations of motion is $R_q=\big(g(z)-g^{[q]}(z)\big)x$.

\begin{proposition}[Energy of the modes]\label{prop:modes}
For $x=A\,e^{\lambda t}+B\,e^{-\lambda t}$ and $L=\sum_{n\le q}\tilde\ell_n$,
the Ostrogradsky energy is, with $z=-\lambda^2$,
\begin{equation}\label{eq:H_modes}
\begin{split}
H^{[q]}={}&-\tfrac12\,g^{[q]}(z)\big(A^2e^{2\lambda t}
+B^2e^{-2\lambda t}\big)\\
&+AB\,\big(2z\,g^{[q]\prime}(z)-g^{[q]}(z)\big).
\end{split}
\end{equation}
\end{proposition}

\begin{demonstration}
For $\tilde\ell_n$, $P^{(m)}=(-1)^{n-m-1}b_nx^{(2n-m-1)}$.
Insert the two exponentials into $H=\sum_mP^{(m)}x^{(m+1)}-\tilde\ell_n$. The
diagonal terms give $b_n\lambda^{2n}e^{\pm2\lambda t}\big[\sum_{m=0}^{n-1}
(-1)^{n-m-1}-\tfrac12\big]=-\tfrac12b_nz^ne^{\pm2\lambda t}$; the cross
terms give $AB\,b_n\lambda^{2n}(-1)^n\big[2n-1\big]=AB\,(2n-1)b_nz^n$.
Summing over
$n$ gives \eqref{eq:H_modes}.
\end{demonstration}

On a solution of the infinite-order equation, $g(z)=0$ with $z\ne0$, and two
consequences follow. (i) The limit $H^{[\infty]}=\lim_qH^{[q]}$ exists
whenever the series $g(z)$ and $zg'(z)$ converge, in particular when $|z|$ is
smaller than the radius of convergence $\rho$ of $g$, and then
\begin{equation}\label{eq:H_infty}
H^{[\infty]}=2AB\,z\,g'(z),
\end{equation}
which is conserved. In this case conditions (i)--(ii) of
Corollary~\ref{cor:Jinfty} hold, since $|R_q|\le\sum_{n>q}|b_n||z|^n|x|$ is
bounded on compact intervals and tends to zero; the corollary is thus
realized in a nontrivial way, with truncated energies that drift according to
\eqref{eq:drift} and a limit that is exactly conserved. (ii) If $|z|>\rho$,
the terms of the series do not tend to zero, $H^{[q]}$ diverges and $R_q$
does not tend to zero. A root of the analytically continued (resummed)
characteristic function then defines a solution of the nonlocal kernel that
is \emph{not} the limit of the derivative expansion.

\begin{figure}[t]
\centering
\includegraphics[width=\columnwidth]{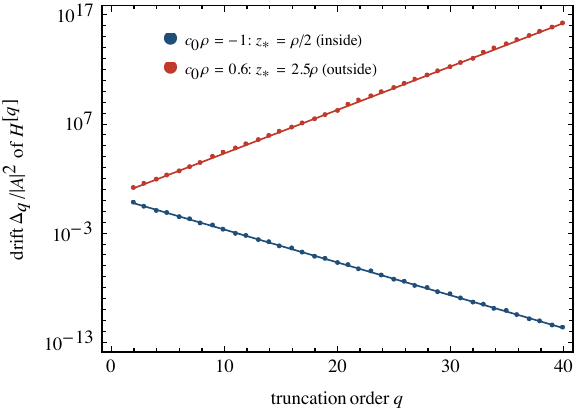}
\caption{Drift $\Delta_q=2|A|^2|g^{[q]}(z_*)|$ of the truncated energy
$H^{[q]}$ on the exact oscillating solution of the small-amplitude sector,
for $f(z)=c_0/(1-z/\rho)$, $m=\rho=1$. Blue: $c_0\rho=-1$, root inside the
disk of convergence; the truncated energies converge to the conserved
$H^{[\infty]}$ (Corollary~\ref{cor:Jinfty}). Red: $c_0\rho=0.6$, root of the
resummed kernel outside the disk; the drift diverges and $J_\infty$ does not
exist.}
\label{fig:drift}
\end{figure}

\emph{Example.} Take $c_k=c_0\rho^{-k}$, i.e.\ $f(z)=c_0/(1-z/\rho)$. Besides
$z=0$ (free motion, $H=\tfrac12mv^2$), the characteristic equation $m+zf(z)=0$
has the single root
\begin{equation}
z_*=\frac{m\rho}{m-c_0\rho}.
\end{equation}
If $c_0<0$, then $0<z_*<\rho$: the root lies inside the disk of convergence,
and $\lambda=\pm i\omega$, $\omega^2=z_*$, is an oscillation whose
conserved energy, for the real solution $B=\bar A$, is
$H^{[\infty]}=2|A|^2\omega^2g'(\omega^2)$ with $g'(z_*)=-m+z_*^2f'(z_*)<0$.
This mode is a ghost, as for $q=2$ (the limit $\rho\to\infty$, where
$g'=-m$). If instead $0<c_0\rho<2m$ with $c_0\rho\neq m$, then
$|z_*|>\rho$: the resummed kernel
has a solution which the series theory does not reproduce, and $J_\infty$
does not exist on it. The convergence conditions of
Corollary~\ref{cor:Jinfty} therefore have physical content: they select the
solutions of the nonlocal theory on which the Noether charges are defined by
the derivative expansion. The time-dependent part of \eqref{eq:H_modes}
makes this visible: for the real solution $B=\bar A$, the truncated energy
oscillates with peak-to-peak drift $\Delta_q=2|A|^2|g^{[q]}(z_*)|$, which is
exactly the tail of the characteristic series at the root. As shown in
Fig.~\ref{fig:drift}, $\Delta_q$ decreases geometrically, as
$(z_*/\rho)^q$, when the root lies inside the disk of convergence and grows in
the same way when it lies outside. 

The field-theory analogue, in which the nonlocal operator is a function of
the d'Alembertian, is discussed in Sec.~\ref{sec:conclusion}.

\begin{remark}
In the case $\rho = \infty$, the disk of convergence is the entire complex plane $\mathbb{C}$, so every exponential mode (every root of $g(z)=0$) automatically satisfies conditions (i)-(ii) of Corollary~\ref{cor:Jinfty}.
\end{remark}

\section{Extension to the full Poincar\'e group}
\label{sec:full}

\subsection{Setting and generators}

We now consider a worldline in $(d+1)$-dimensional Minkowski space, in the time
gauge, described by $d$ functions $x^i(t)$, $i=1,\dots,d$, with $t=x^0$
and metric $g=\mathrm{diag}(+,-,\dots,-)$. An infinitesimal Lorentz
transformation $x^\mu\to x^\mu+\epsilon^\mu{}_\nu x^\nu$, with
$\epsilon_{\mu\nu}=-\epsilon_{\nu\mu}$, depends on $d(d+1)/2$ parameters:
$d$ boost parameters $\beta_i:=\epsilon_{0i}$ and $d(d-1)/2$ rotation
parameters $\omega_{ij}:=-\epsilon_{ij}=-\omega_{ji}$. In components,
\begin{equation}\label{eq:lorentz_d}
\bar t=t+\beta_j x^j,\qquad
\bar x^i=x^i+\beta_i\,t+\omega_{ij}\,x^j ,
\end{equation}
which preserves $t^2-x^ix^i$ and reduces to \eqref{eq:boost} for $d=1$,
$\beta_1=\epsilon$. We also include translations $\delta t=h^0$,
$\delta x^i=h^i$, so that the whole Poincar\'e group is treated at once. A
generic generator is
\begin{equation}\label{eq:Xd}
X=\xi\,\partial_t+\eta^i\,\partial_{x^i},\qquad
Q^i:=\eta^i-\xi\,x'^i ,
\end{equation}
with, for the Lorentz part,
\begin{equation}\label{eq:xi_eta_d}
\xi=\beta_jx^j,\qquad
\eta^i=\beta_it+\omega_{ij}x^j,
\end{equation}
and, for translations, $\xi=h^0$, $\eta^i=h^i$. The prolongation $X^{(q)}$
and $\eta^{i(n)}$ are defined as in \eqref{eq:Xq_def}, component by component.

\begin{lemma}\label{lem:prolongation_d}
For any generator \eqref{eq:Xd} with $\xi,\eta^i$ depending on $(t,x^j)$,
$\eta^{i(n)}=\D^nQ^i+\xi\,x^{i(n+1)}$ for all $n\ge0$. For the Lorentz
transformation \eqref{eq:lorentz_d},
\begin{align}
\eta^{i(1)}&=\beta_i+\omega_{ij}x'^j-\beta_jx'^jx'^i,\\
\eta^{i(n)}&=\omega_{ij}x^{j(n)}\nonumber\\
&\quad-\beta_j\sum_{m=1}^{n}\binom nm x^{j(m)}x^{i(n+1-m)}\quad(n\ge2).
\end{align}
\end{lemma}

\begin{demonstration}
The first statement follows by the same induction as in
Lemma~\ref{lem:prolongation}, using $\eta^{i(0)}=Q^i+\xi x'^i$. For the
Lorentz generator, $Q^i=\beta_it+\omega_{ij}x^j-\beta_jx^jx'^i$; for $n=1$,
$\D Q^i+\xi x''^i=\beta_i+\omega_{ij}x'^j-\beta_jx'^jx'^i$, and for $n\ge2$ the
Leibniz rule applied to $\D^n(x^jx'^i)$ gives the result, the term $m=0$
cancelling against $\xi\,x^{i(n+1)}$. For $d=1$ and $\omega=0$ one recovers
\eqref{eq:etan}.
\end{demonstration}

\subsection{Noether identity and Poincar\'e charges}

Let $L=L(x^i,x'^i,\dots,x^{i(q)})$ have no explicit $t$-dependence, and set
$L^{(n)}_i=\partial L/\partial x^{i(n)}$,
\begin{align}\label{eq:EL_d}
E_i(L)&=\sum_{n=0}^{q}(-1)^n\D^nL_i^{(n)},\\
P_i^{(m)}&=\sum_{n=m+1}^{q}(-1)^{n-m-1}\D^{\,n-m-1}L_i^{(n)} .
\end{align}

\begin{theorem}[Poincar\'e Noether identity]\label{thm:noether_d}
For every generator \eqref{eq:Xd}, define
\begin{equation}\label{eq:J_d}
J=\xi\,L+\sum_{m=0}^{q-1}\big(\D^mQ^i\big)\,P_i^{(m)}.
\end{equation}
Then, for every smooth trajectory,
\begin{equation}\label{eq:identity_d}
\D J=X^{(q)}L+(\D\xi)\,L-Q^i\,E_i(L).
\end{equation}
For Lorentz transformations, $\D\xi=\beta_jx'^j$.
\end{theorem}

\begin{demonstration}
By Lemma~\ref{lem:prolongation_d} and $\partial_t L=0$,
$X^{(q)}L=\sum_{n=0}^q(\D^nQ^i)L^{(n)}_i+\xi\,\D L$, hence
$X^{(q)}L+(\D\xi)L=\sum_{n}(\D^nQ^i)L_i^{(n)}+\D(\xi L)$. Applying
\eqref{eq:ibp} to each component with $f=Q^i$, $g=L_i^{(n)}$ and relabeling
$m=n-1-j$ gives \eqref{eq:identity_d}, with the current \eqref{eq:J_d}.
\end{demonstration}

For $d=1$, $\beta_1=1$ and $\omega=0$, Theorem~\ref{thm:noether_d} reduces to
Theorem~\ref{thm:noether}. The identity \eqref{eq:identity_d} is linear in the
parameters $(h^0,h^i,\beta_i,\omega_{ij})$, so each parameter yields a
separate current.

\begin{corollary}[Poincar\'e charges]\label{cor:poincare}
Define the Ostrogradsky Hamiltonian
$H=\sum_{m=0}^{q-1}P_i^{(m)}x^{i(m+1)}-L$ and
$\mathcal L_{ij}=\sum_{m=0}^{q-1}\big(x^{i(m)}P_j^{(m)}-x^{j(m)}P_i^{(m)}\big)$.
The currents \eqref{eq:J_d} associated with the generators of the Poincar\'e
group are the following.
\begin{itemize}
\item Time translation ($\xi=1$, $\eta^i=0$): $J=-H$. Since $L$ is autonomous,
$X^{(q)}L=0$ and $\D H=-x'^iE_i(L)$, so $H$ is conserved on shell.
\item Spatial translation ($\xi=0$, $\eta^i=h^i$): $J=h^iP_i^{(0)}$, conserved
on shell if $\partial L/\partial x^i=0$.
\item Rotations ($\xi=0$, $\eta^i=\omega_{ij}x^j$):
$J=-\tfrac12\,\omega_{ij}\mathcal L_{ij}$.
\item Boosts: $J=\beta_kK^k$, with
\begin{equation}\label{eq:K}
K^k=t\,P^{(0)}_k-x^kH+N^k,
\end{equation}
\begin{equation}
N^k=P_k^{(1)}-\sum_{m=1}^{q-1}\sum_{l=1}^{m}\binom ml\,
x^{k(l)}x^{i(m+1-l)}P_i^{(m)}.
\end{equation}
\end{itemize}
For $q=1$, $N^k=0$ and $K^k=t\,p_k-x^kH$ is the usual boost charge. For
$q\ge2$ the boost charge contains the additional term $N^k$.
\end{corollary}

\begin{demonstration}
Insert $\xi,Q^i$ into \eqref{eq:J_d}. For the boost, $\D Q^i=\delta^{ik}-
\D(x^kx'^i)$ and $\D^mQ^i=-\D^m(x^kx'^i)$ for $m\ge2$. The terms
$x^kL$ and $-x^kx^{i(m+1)}P_i^{(m)}$ (the Leibniz term $l=0$) combine into
$-x^kH$, the term $t\delta^{ik}P_i^{(0)}$ gives $t P_k^{(0)}$, the
constant in $\D Q^i$ gives $P_k^{(1)}$, and the remaining Leibniz terms
$l\ge1$ give the sum in $N^k$. The rotation part follows from
$\omega_{ij}x^{j(m)}P_i^{(m)}=-\frac12\omega_{ij}(x^{i(m)}P_j^{(m)}
-x^{j(m)}P_i^{(m)})$.
\end{demonstration}

\paragraph*{Interpretation: $N^k$ is an internal spin.}
Define the momenta $p^0:=H$, $p^k:=P_k^{(0)}$ and the antisymmetric tensor
\begin{align}
S^{ij}&:=\sum_{m=1}^{q-1}\big(x^{i(m)}P_j^{(m)}-x^{j(m)}P_i^{(m)}\big),
\nonumber\\
S^{0k}&:=N^k. \label{eq:spin}
\end{align}
Then the charges of Corollary~\ref{cor:poincare} take the form
\begin{equation}
\mathcal L_{ij}=x^ip_j-x^jp_i+S^{ij},\qquad
K^k=t p^k-x^kH+S^{0k},
\end{equation}
i.e.\ $M^{\mu\nu}=x^\mu p^\nu-x^\nu p^\mu+S^{\mu\nu}$, which is precisely the
structure of the angular momentum of a spinning particle. The tensor
$S^{\mu\nu}$ is built only from the derivatives $x^{i(m)}$, $m\ge1$, and from
the momenta, so it is invariant under spacetime translations: it is an
\emph{internal} angular momentum, generated by the higher derivatives and
absent for $q=1$. This is the time-gauge counterpart of the spin found in the
covariant treatment of curvature-dependent Lagrangians
\cite{Nesterenko1995,Plyushchay1988,Plyushchay1989,KuznetsovPlyushchay1993}.
The spin is not separately conserved. On shell, using $\D H=0$ and
$\D p^k=\partial L/\partial x^k=0$,
\begin{equation}\label{eq:zitter}
\D S^{0k}=x'^kH-p^k ,
\end{equation}
so $S^{0k}$ is conserved if and only if the velocity equals $p^k/H$.
Similarly, the conservation of $\mathcal L_{ij}$ gives
$\D S^{ij}=-(x'^ip^j-x'^jp^i)$ on shell, so that $S^{ij}$ is conserved when
the velocity is parallel to the momentum. For
$q\ge2$ the velocity and the momentum are in general not aligned, the
classical analogue of the Zitterbewegung of spinning particles, made precise
in Sec.~\ref{sec:mass}. Section~\ref{sec:mass} also shows that, in $1+1$
dimensions, the jet-space action of the boost on the charges reproduces the
Poincar\'e algebra; the corresponding Poisson brackets in the Ostrogradsky
phase space are not studied here. 

\begin{remark}
One might worry that passing to the Ostrogradsky canonical variables $(q_m, P^{(m)})$ introduces subtleties, in particular for degenerate Lagrangians. Here, the nondegeneracy of $\partial^2 L / \partial (x^{(q)})^2$ guarantees the isomorphism between the jet-space action and the Hamiltonian action via Poisson brackets.
\end{remark}

\subsection{Invariant Lagrangians}

\begin{corollary}\label{cor:invariant_d}
Suppose $L$ satisfies $X^{(q)}L+(\D\xi)L=0$ for all boosts and rotations.
Then
\begin{enumerate}
\item[(a)] $\partial L/\partial x^i=0$ for all $i$, since $t$ appears
explicitly only through the term $\beta_it\,\partial L/\partial x^i$;
\item[(b)] $L$ is invariant under simultaneous rotations of the vectors
$x^{(n)}$, $n\ge1$; by the first fundamental theorem of invariant theory for
$SO(d)$, it is a function of their scalar products and, when $q\ge d$, of the
determinants formed from $d$ of them;
\item[(c)] the boost condition reduces to
\begin{equation}
\sum_{i}\sum_{n=1}^{q}\eta^{i(n)}_{(k)}\frac{\partial L}{\partial x^{i(n)}}
+x'^kL=0,
\end{equation}
where $\eta^{i(n)}_{(k)}$ is the coefficient of $\beta_k$ in
Lemma~\ref{lem:prolongation_d}.
\end{enumerate}
\end{corollary}

\begin{example}[Arbitrary $d$]
For $q=1$ and $L=L(u)$ with $u=x'^ix'^i$, condition (c) reads
$2x'^k(1-u)\,L'(u)+x'^kL=0$, hence
\begin{equation}
L=-m\sqrt{1-u}
\end{equation}
again. For $q=2$, let $\gamma=(1-u)^{-1/2}$ and
\begin{equation}
\alpha^2=\gamma^4\,x''^ix''^i+\gamma^6\,(x'^ix''^i)^2,
\end{equation}
the squared proper acceleration, $\alpha^2=-a_\mu a^\mu$ with
$a^\mu=\dd^2x^\mu/\dd\sigma^2$. Then
\begin{equation}
L=\sqrt{1-u}\,f(\alpha^2)
\end{equation}
is boost and rotation invariant for any $f$. For $d=1$, $\alpha^2=a^2$ and it
reduces to the Lagrangian of Sec.~\ref{sec:invariance} with $f$ even in $a$;
odd functions of $a$ are allowed only for $d=1$, where there are no
rotations. The invariance follows without computation: $\sqrt{1-u}\,\dd t
=\dd\sigma$ and $\alpha^2=-g_{\mu\nu}a^\mu a^\nu$ is a Lorentz scalar, so
$L\,\dd t=f(\alpha^2)\,\dd\sigma$ is invariant under the whole Lorentz
group.
\end{example}

\section{Poincar\'e algebra, mass and spin in $1+1$ dimensions}
\label{sec:mass}

We return to $1+1$ dimensions, where the Poincar\'e charges reduce to three, $H$,
$p:=P^{(0)}$ and $K=tp-xH+S^{01}$, and show that they carry a complete and
explicit realization of the Poincar\'e algebra. Throughout, $L$ satisfies
\eqref{eq:invariance}, so that by Theorem~\ref{thm:general_invariant}
$L=s\,F(a,a_1,\dots,a_r)$ with $r=q-2$ (for $q=1$, $F=-m$), and we use the covariant quantities
$\mathcal E$ and $\mathcal H$ of Eq.~\eqref{eq:pi_H}.

\begin{proposition}[Lorentz action]\label{prop:algebra}
Identically on the jet space,
\begin{equation}\label{eq:algebra}
X^{(2q-1)}p=H,\quad X^{(2q-1)}H=p,\quad X^{(2q-1)}S^{01}=0 ,
\end{equation}
while the translations act by $\partial_tK=p$, $\partial_xK=-H$ and leave $H$,
$p$, $S^{01}$ invariant. Consequently the mass
\begin{equation}
M^2:=H^2-p^2
\end{equation}
is invariant under the whole Poincar\'e group.
\end{proposition}

\begin{demonstration}
$H$, $p$ and $S^{01}$ are built from $L$, its partial derivatives and their
total derivatives; since $L$ depends neither on $t$ nor on $x$, neither do
they, and neither do $X^{(2q-1)}H$, $X^{(2q-1)}p$, $X^{(2q-1)}S^{01}$, because
the components $\eta^{(n)}$, $n\ge1$, do not. Apply $X^{(2q-1)}$ to
$K=tp-xH+S^{01}$ and use Proposition~\ref{prop:XJ}:
$0=xp+t\,Xp-tH-x\,XH+XS^{01}$. The coefficients of $t$, of $x$ and the
remainder must vanish separately, which gives \eqref{eq:algebra}. The
translation statements are immediate, and $X(H^2-p^2)=2Hp-2pH=0$.
\end{demonstration}

Equations \eqref{eq:algebra} are the infinitesimal Lorentz transformation of
the two-vector $(H,p)$ and the invariance of the boost charge under its own
boost: the charges transform under the jet-space action exactly as the
generators of the Poincar\'e algebra, with no central term. By
Theorem~\ref{thm:general_invariant} (applied to functions of weight zero),
$M^2$ is a function of $a,a_1,\dots$ only. The next result makes all the
charges explicit.

\begin{theorem}[Momentum and spin]\label{thm:momentum}
With $u^\mu=\gamma(1,v)$ and $n^\mu=\gamma(v,1)$ the Frenet frame of the
worldline,
\begin{equation}\label{eq:pmu}
p^\mu\equiv(H,p)=\big(a\,\mathcal E+\mathcal H\big)\,u^\mu
-\dot{\mathcal E}\,n^\mu ,\qquad S^{01}=\mathcal E ,
\end{equation}
so that $M^{\mu\nu}=x^\mu p^\nu-x^\nu p^\mu+\mathcal E\,(u^\mu n^\nu-u^\nu
n^\mu)$, and
\begin{equation}\label{eq:mass}
M^2=\big(a\,\mathcal E+\mathcal H\big)^2-\dot{\mathcal E}^{\,2} .
\end{equation}
\end{theorem}

\begin{demonstration}
Call $p_c^\mu$ the right-hand side of \eqref{eq:pmu}. With the Frenet
equations $\dot u=a\,n$, $\dot n=a\,u$,
\begin{equation}
\dot p_c^{\,\mu}=\big(\dot a\,\mathcal E+\dot{\mathcal H}\big)u^\mu
+\big(a^2\mathcal E+a\mathcal H-\ddot{\mathcal E}\big)n^\mu=-E(L)\,n^\mu ,
\end{equation}
where we used the energy identity $\dot{\mathcal H}=-\dot a\,\mathcal E$ of
the Lagrangian $F(a,\dot a,\dots)$ and Proposition~\ref{prop:cov_eom}. Since
$\D=s\,\dd/\dd\sigma$ and $s\,n^\mu=(v,1)$, this gives $\D p_c^\mu=-E\,(v,1)$.
On the other hand, the Noether identities of Corollary~\ref{cor:poincare}
give $\D H=-vE$ and $\D p=L_0-E=-E$ off shell. Hence
$\D(H-p_c^0)=\D(p-p_c^1)=0$ identically, and both differences are constants
(Step~2 of Proposition~\ref{prop:XJ}). Similarly, $K_c:=tp-xH+\mathcal E$
satisfies $\D K_c=p-vH-QE+s\dot{\mathcal E}=-QE=\D K$, because
$p-vH=-s\dot{\mathcal E}$ by \eqref{eq:pmu}. The three constants are fixed at
the rest jet $j_0$ ($t=x=0$, $x^{(k)}=0$ for $k\ge1$). Under $t\to\lambda t$,
$x\to\lambda x$, $a_k$ has weight $-(k+1)$, so each of its monomials contains
derivatives $x^{(j)}$, $j\ge2$, of total weight $k+1$: hence $a_k$ and
$\partial a_k/\partial x'$ vanish at $j_0$, and $\partial a_k/\partial
x''|_{j_0}=\delta_{k0}$. Moreover every total derivative of a
$t$-independent function vanishes at $j_0$. Therefore $H(j_0)=-L(j_0)=-F(0)
=p_c^0(j_0)$, $p(j_0)=L_1(j_0)=0=p_c^1(j_0)$ and $K(j_0)=N(j_0)=L_2(j_0)=
F_a(0)=\mathcal E(j_0)=K_c(j_0)$. Finally $M^2=p_c\cdot p_c$ with $u\cdot u=1$,
$n\cdot n=-1$, $u\cdot n=0$.
\end{demonstration}

For $r=0$, Eqs.~\eqref{eq:pmu}--\eqref{eq:mass} give
$M^2=(af'-f)^2-(f''\dot a)^2$, in agreement with \eqref{eq:H_sf}, and
$S^{01}=f'(a)$, which reproduces $K=c_0a_*$ on the self-accelerated
solutions, now recognized as \emph{massless}, $M=0$. In the covariant
formalism, the mass and spin of curvature-dependent particles are known to be
simple functions of the Frenet--Serret curvatures
\cite{Nesterenko1995,Arreaga2001}; Theorem~\ref{thm:momentum} recovers these
expressions in $1+1$ dimensions, for Lagrangians of arbitrary order, and its
content is the identification of the time-gauge Ostrogradsky charges $H$,
$p$ and $N$ with the covariant objects, in particular $N=S^{01}=\mathcal E$. 

\begin{corollary}[Center of energy]\label{cor:center}
If $H\neq0$, the point $x_c(t):=(tp-K)/H$ moves uniformly with velocity
$p/H$, and
\begin{equation}\label{eq:zb}
x-x_c=\frac{\mathcal E}{H}.
\end{equation}
If $M^2>0$, in the rest frame of $p^\mu$ the worldline is given by
\begin{equation}\label{eq:quadrature}
\frac{\dd t}{\dd\sigma}=\frac{a\mathcal E+\mathcal H}{M},\qquad
x-x_c=\frac{\mathcal E}{M},
\end{equation}
so that every solution is determined by the single function $a(\sigma)$. For
$r=0$, $a(\sigma)$ follows by quadrature from the first integral
\eqref{eq:mass}, $f''(a)^2\,\dot a^2=(af'-f)^2-M^2$, and the model is
integrable, as known from the covariant analysis
\cite{Nesterenko1995,Arreaga2001}.
\end{corollary}

\begin{demonstration}
$\D x_c=(p-\D K)/H-(tp-K)\D H/H^2=p/H$ on shell. Equation \eqref{eq:zb} is
$K=tp-xH+S^{01}$ with $S^{01}=\mathcal E$. In the rest frame, $p^1=0$ in
\eqref{eq:pmu} gives $\gamma v(a\mathcal E+\mathcal H)=\gamma\dot{\mathcal E}$
and $H=M$ gives $\gamma=(a\mathcal E+\mathcal H)/M$.
\end{demonstration}

\begin{figure*}[t]
\centering
\includegraphics[width=\textwidth]{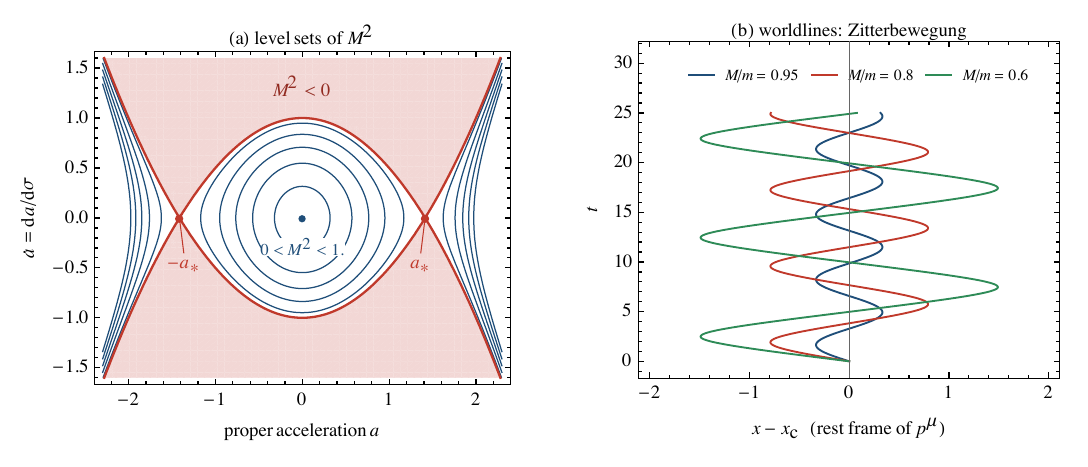}
\caption{Model $f(a)=-m+\tfrac12c_0a^2$ with $m=1$, $c_0=-1$. (a) Level sets
of the Poincar\'e invariant $M^2=(m+\tfrac12c_0a^2)^2-c_0^2\dot a^2$
[Eq.~\eqref{eq:mass}] in the plane $(a,\dot a)$. Closed orbits have
$0<M^2<m^2$; the separatrix $M=0$ (red) connects the self-accelerated
solutions $\pm a_*$, which are saddle points; the shaded region has $M^2<0$.
(b) Worldlines in the rest frame of $p^\mu$, obtained from
Eq.~\eqref{eq:quadrature}: the particle oscillates around the center of
energy $x_c$ with displacement $\mathcal E/M=c_0a/M$, and the amplitude grows
as $M$ decreases.}
\label{fig:phase}
\end{figure*}

Equation \eqref{eq:zb} is the precise form of the Zitterbewegung announced in
Sec.~\ref{sec:full}: the particle oscillates around a uniformly moving center
of energy, with a displacement equal to spin over energy. For the model
\eqref{eq:model} truncated at $q=2$, $f=-m+\tfrac12c_0a^2$ with $c_0<0$, the
first integral reads $c_0^2\dot a^2=(m+\tfrac12c_0a^2)^2-M^2$, whose level
sets are shown in Fig.~\ref{fig:phase}(a). For $0<M<m$ the proper
acceleration oscillates between $\pm a_-$, $a_-^2=2(m-M)/|c_0|$, and by
\eqref{eq:quadrature} the particle oscillates around $x_c$ with amplitude
$|c_0|a_-/M$ [Fig.~\ref{fig:phase}(b)]. For small amplitudes the frequency is
$\omega^2=m/|c_0|$, which is the root $z_*$ of Sec.~\ref{sec:linear} for
$q=2$, and $M<m$: the oscillation lowers the energy, in agreement with its
ghost character found there. The separatrix $M=0$ ends on the
self-accelerated solutions $a=\pm a_*$, which appear as saddle points,
consistently with the instability found in Sec.~\ref{sec:model}. Outside the
separatrix, in the shaded region of Fig.~\ref{fig:phase}(a), $M^2<0$: the
momentum is spacelike, a tachyonic sector of the classical rigid particle.
The remaining orbits, with $|a|>a_*$, have $p\cdot u=af'-f<0$, i.e.\ a
momentum pointing to the past with respect to the four-velocity.

\section{Conclusion and outlook}
\label{sec:conclusion}

We have computed the prolongation of Poincar\'e transformations to arbitrary
order in the derivatives, proved the Noether identities
\eqref{eq:identity} and \eqref{eq:identity_d}, and obtained explicit conserved
currents in terms of Ostrogradsky momenta. The angular momentum has the
structure $x^\mu p^\nu-x^\nu p^\mu+S^{\mu\nu}$ of a spinning particle, the
internal spin $S^{\mu\nu}$ being generated by the higher derivatives; its
boost component is the term $N^k$ by which the boost charge differs from
$t p-xH$, and its non-conservation, Eq.~\eqref{eq:zitter}, expresses the
misalignment of velocity and momentum. The boost current is invariant under
the boost it generates only when the boost acts on the full Ostrogradsky
state, i.e.\ is prolonged to order $2q-1$ (Proposition~\ref{prop:XJ}); at
first order, invariance of $L\,\dd t$, conservation of $J_1$ and invariance
of $J_1$ coincide and single out the free particle. In $(1+1)$ dimensions the
invariant Lagrangians of order $q$ are $\sqrt{1-x'^2}\,F(a,\dot a,\dots,
a^{(q-2)})$, and in $d$ dimensions they include $\sqrt{1-u}\,f(\alpha^2)$ at
$q=2$. For
infinite-derivative Lagrangians, the conservation defect of the truncated
currents is the tail of the equations of motion weighted by the
characteristic, which gives checkable conditions for the exact conservation
of $J_\infty$. In an explicit nonlocal model these conditions hold trivially
on trajectories of constant acceleration, where the model has exact,
zero-energy self-accelerated solutions, which the covariant equations of
motion show to be linearly unstable for every nonlocal completion, and
nontrivially in the small-amplitude sector, where they select the modes
lying inside the disk of convergence of the characteristic function. Finally,
in $1+1$ dimensions the charges realize the Poincar\'e algebra without
central term, the momentum, spin and mass take the closed covariant form of
Theorem~\ref{thm:momentum}, the particle oscillates around a uniformly moving
center of energy with displacement $\mathcal E/H$, and the integrability of
the second-order model \cite{Nesterenko1995,Arreaga2001} is recovered; its
phase portrait exhibits bounded Zitterbewegung, the massless
self-accelerated saddles and a tachyonic sector.

Several directions deserve further investigation. First, the convergence
conditions of Corollary~\ref{cor:Jinfty} should be studied beyond the
small-amplitude sector, in particular for the boost charge, and including the
nonlinear corrections to the modes of Sec.~\ref{sec:linear}, which requires a proper formulation of the initial-value problem
\cite{BarnabyKamran2008,CarlssonPradoReyes2016,Heredia2023}. Second, the jet-space realization of the Poincar\'e algebra of
Proposition~\ref{prop:algebra} should be extended to $d$ dimensions and
related to the Poisson algebra of the charges in the Ostrogradsky phase
space; Theorem~\ref{thm:general_invariant} should likewise be extended to
$d$ spatial dimensions in
terms of Frenet--Serret curvatures \cite{Nesterenko1995,Arreaga2001}. Third,
the connection with the Fokker-type action \eqref{eq:fokker_action}, whose
symmetries were studied in Ref.~\cite{GaidaTretyak1996}, should be made
explicit.

\paragraph*{Outlook: field theories.}
The strategy extends naturally to a scalar field with the nonlocal action
\begin{equation}\label{eq:F}
\begin{split}
S[\phi]&=\int\dd^Dx\Big[\tfrac12\,\phi\,F(\Box)\,\phi-V(\phi)\Big],\\
F(\Box)&=\sum_{n\ge0}w_n\Box^n,
\end{split}
\end{equation}
with $\Box=g^{\mu\nu}\partial_\mu\partial_\nu$. Since $\Box$ is a Lorentz
scalar, every truncation $F_q(\Box)=\sum_{n\le q}w_n\Box^n$ is a local,
manifestly Lorentz-invariant higher-derivative theory, for which the Noether
currents are given by the standard formulas for higher-derivative fields
\cite{Olver1993,KrivoruchenkoTursunov2016}; at first order they reduce to
$J^\mu=-\xi_\nu T^{\mu\nu}$, with $T^{\mu\nu}$ the canonical energy--momentum
tensor. By linearity, the currents of the truncations are partial sums of a
series in the $w_n$, and the argument of Theorem~\ref{thm:defect} carries
over: the divergence of each partial sum is the characteristic times the tail
of the field equations. The worldline results suggest that the convergence of
the limit will select a class of solutions determined by the analytic
structure of $F$, as for the entire functions of ghost-free
infinite-derivative gravity \cite{Biswas2012,Modesto2012}. Working this out
is the natural continuation of this paper.

\section*{Acknowledgments}
V.L. warmly thanks the little crab and the swan of destiny.

D.O.S. expresses his sincere gratitude to all those who supported him during his unjust imprisonment at Ouidah Prison, where much of this work was carried out, and to all those who tirelessly fought for his release. Their solidarity and unwavering support will always be remembered.

\section*{Data availability}

No experimental data were created or analyzed in this study.  The figures are generated by a very simple Mathematica code that will be available in the GitHub repository after the article is published.

\appendix

\section{Direct proof of Corollary~\ref{cor:conservation}}
\label{app:direct}

\begin{demonstration}[Direct proof of Corollary~\ref{cor:conservation}]
We differentiate the explicit form \eqref{eq:Jq} term by term; this provides
an independent check of Theorem~\ref{thm:noether}. We use $\D Q=1-\D(xx')=1-x'^2-xx''$ and
$\D L=\sum_{k=0}^{q}x^{(k+1)}L_k$.

\emph{Step 1 (differentiation).} The first two terms give
\begin{align}
\D(xL)+\D(QL_1)={}&x'L+\sum_{k=0}^{q}x\,x^{(k+1)}L_k\nonumber\\
&+(\D Q)L_1+Q\,\D L_1 .
\end{align}
For each $n\ge2$, the bracket in \eqref{eq:Jq} gives
\begin{equation}
\begin{split}
&(-1)^{n-1}(\D Q)\,\D^{n-1}L_n+(-1)^{n-1}Q\,\D^nL_n\\
&\quad+(-1)^n\D^{n-1}L_n+A_n+B_n,
\end{split}
\end{equation}
where
\begin{align}
A_n&=\sum_{k=1}^{n-1}(-1)^k\D^{n+1-k}(xx')\,\D^{k-1}L_n,\nonumber\\
B_n&=\sum_{k=1}^{n-1}(-1)^k\D^{n-k}(xx')\,\D^{k}L_n.\nonumber
\end{align}

\emph{Step 2 (telescoping).} Setting $j=k-1$ in $A_n$,
$A_n=-\D^n(xx')\,L_n+\sum_{j=1}^{n-2}(-1)^{j+1}\D^{n-j}(xx')\,\D^jL_n$,
while
$B_n=\sum_{k=1}^{n-2}(-1)^k\D^{n-k}(xx')\,\D^kL_n
+(-1)^{n-1}\D(xx')\,\D^{n-1}L_n$. The two sums cancel, and
\begin{equation}
A_n+B_n=-\D^n(xx')\,L_n+(-1)^{n-1}\D(xx')\,\D^{n-1}L_n .
\end{equation}

\emph{Step 3 (cancellation of $\D^{n-1}L_n$).} The coefficient of
$\D^{n-1}L_n$ is now $(-1)^{n-1}\big[\D Q+\D(xx')\big]+(-1)^n
=(-1)^{n-1}+(-1)^n=0$.

\emph{Step 4 (Euler--Lagrange expression).} The terms proportional to $Q$
combine into $Q\big[\D L_1-\sum_{n\ge2}(-1)^n\D^nL_n\big]=Q\,(L_0-E(L))$, by
\eqref{eq:EL}.

\emph{Step 5 (prolongation).} Collecting the remaining terms by $L_n$:
$Q L_0+xx'L_0=t L_0$;
$(\D Q)L_1+xx''L_1=(1-x'^2)L_1=\eta^{(1)}L_1$; and, for $n\ge2$, by the
Leibniz rule,
$x\,x^{(n+1)}L_n-\D^n(xx')L_n=-\sum_{k=1}^{n}\binom nk x^{(k)}x^{(n+1-k)}L_n
=\eta^{(n)}L_n$. Since $\partial_t L=0$, this yields
\begin{align}
\D J_q&=x'L+t\frac{\partial L}{\partial x}+(1-x'^2)\frac{\partial L}{\partial
x'}\nonumber\\
&\quad-\sum_{n=2}^{q}\sum_{k=1}^{n}\binom nk x^{(k)}x^{(n+1-k)}
\frac{\partial L}{\partial x^{(n)}}-Q\,E(L)\nonumber\\
&=X^{(q)}L+x'L-Q\,E(L),
\end{align}
which is \eqref{eq:identity}. If \eqref{eq:invariance} holds, the first two
terms cancel and $\D J_q=-Q\,E(L)$, which vanishes on every solution of
$E(L)=0$. Conversely, on a solution $\D J_q=X^{(q)}L+x'L$, so the conservation
of $J_q$ on all solutions implies that $X^{(q)}L+x'L$ vanishes on them.
\end{demonstration}

\section{Proof of Proposition~\ref{prop:cov_eom}}
\label{app:cov}

Parametrize the worldline by its proper time $\sigma$, with unit tangent $u$
($u\cdot u=1$) and unit normal $n$ ($n\cdot n=-1$, $u\cdot n=0$). In $1+1$
dimensions the Frenet equations read $\dd u/\dd\sigma=a\,n$ and
$\dd n/\dd\sigma=a\,u$. Consider the normal variation $\delta x^\mu=\varphi\,
n^\mu$; tangential variations are reparametrizations and do not change an
invariant action. Using the Frenet equations,
$\delta(\dd x/\dd\sigma)=\dot\varphi\,n+a\varphi\,u$, hence
$\delta(\dd\sigma)=a\varphi\,\dd\sigma$ and
$\delta(\dd/\dd\sigma)=-a\varphi\,\dd/\dd\sigma$. Then $\delta u=\dot\varphi\,n$
and, from $a=-n\cdot\dd u/\dd\sigma$ and $n\cdot\delta n=0$,
\begin{equation}
\delta a=\ddot\varphi-a^2\varphi ,\qquad
\delta a_k=\frac{\dd}{\dd\sigma}\delta a_{k-1}-a\varphi\,a_k .
\end{equation}
Unfolding the recursion, $\delta a_k=\partial^k\delta a-\sum_{j=1}^k
\partial^{k-j}(a\varphi\,a_j)$, with $\partial=\dd/\dd\sigma$. The variation of
$S=\int F\,\dd\sigma$ is
$\delta S=\int\big[\sum_kF_{a_k}\delta a_k+a\varphi F\big]\dd\sigma$.
Integrating by parts, the first part of $\delta a_k$ gives
$\int\mathcal E\,\delta a\,\dd\sigma$, and the second gives
$-\int a\varphi\sum_{j\ge1}a_j\sum_{k\ge j}(-\partial)^{k-j}F_{a_k}\,\dd\sigma
=-\int a\varphi\sum_{j\ge1}a_j\pi_{j-1}\,\dd\sigma$. Hence, with one more
integration by parts on $\int\mathcal E(\ddot\varphi-a^2\varphi)$,
\begin{equation}
\delta S=\int\varphi\,\big[\ddot{\mathcal E}-a^2\mathcal E-a\,\mathcal H\big]
\,\dd\sigma .
\end{equation}
Finally, relate this to the time gauge. With $u=\gamma(1,v)$ and
$n=\gamma(v,1)$, the variation $\delta x^\mu=\varphi n^\mu$ changes the
function $x(t)$ at fixed $t$ by $\delta x|_t=\delta x^1-x'\delta x^0
=\gamma(1-v^2)\varphi=s\,\varphi$, so that
$\delta S=\int E(L)\,\delta x|_t\,\dd t=\int E(L)\,\varphi\,\dd\sigma$.
Comparing the two expressions for arbitrary $\varphi$ gives
\eqref{eq:cov_eom}. \hfill$\blacksquare$

\end{document}